\documentstyle[11pt]{article}

          \def\dt{\cal}
          \def\dA{{\dt A}}
          \def\dB{{\dt B}}

          \def\dM{{\dt M}}

          \def\B{{\cal B}}
          \def\C{{\cal C}}

          \def\H{{\cal H}}

          \def\K{{\cal K}}
          
          \def\M{{\cal M}}
          
          \def\O{{\cal O}}
          \def\P{{\cal P}}

          \def\gD{\Delta}

          \def\gO{\Omega}

          \def\gg{\gamma}

          \def\gn{\nu}

          \def\gs{\sigma}

          \def\aloc{\dA_{\rm loc}}
          \def\aqloc{\widetilde{\dA}}

          \def\complex{{\bf C}}

          \def\fpsi{f_{\psia,\psib,A,B}}

          \def\halmos{\quad\begin{flushright}$\Box$\end{flushright}}
          \def\Halmos{\quad\begin{flushright}$\Box$\end{flushright}}
          \def\id{\rm id}

          \def\jt{\tilde \jmath}

          \def\modopt{\gD^{it}}
          \def\modopmt{\gD^{-it}}
          \def\naturals{{\bf N}}
          \def\oa{\O_{1}}
          
          \def\ob{\O_{2}}

          \def\onu{\O_{\gn}}
          \def\ot{\tilde\O}
          \def\poin{\P_+^{\uparrow}}
          \def\pct{P$_1$CT}
          \def\psia{\psi_1}
          \def\psib{\psi_2}
          
          \def\Rd{\reals^{1+s}}

          \def\reals{{\bf R}}

          \def\upg{\widetilde{\P_{+}^{\uparrow}}} 

          \def\Vq{\overline{V}}
          \def\Vpq{\overline{V}_+}

          \def\Vpq{\overline{V}_+}
          
          \def\xv{\vec x}
          \def\yv{\vec y}
\title{Borchers' Commutation Relations\\ and Modular Symmetries}
\author{Bernd Kuckert\thanks{
Universit\"at Hamburg, II. Institut f\"ur Theoretische
Physik, Luruper Chaussee 149, 22761 Hamburg, e-mail:
kuckert@x4u2.desy.de}}
\begin{document}
\frenchspacing
\maketitle

\begin{abstract}
Recently Borchers has shown that in a theory of local observables,
certain unitary and antiunitary operators, which are obtained from
an elementary construction suggested by Bisognano and Wichmann,
commute with the translation operators like
Lorentz boosts and \pct-operators, respectively. We conclude from this
that as soon as the operators considered implement {\em any}
symmetry, this symmetry can be fixed up to at most some translation.
As a symmetry, we admit any unitary or antiunitary operator under
whose adjoint action any algebra of local observables is mapped
onto an algebra which can be localized somewhere in Minkowski space.
\end{abstract}

\section*{0\, Introduction}
It is a classical result due to Bisognano and Wichmann that
in the traditional (Wightman) theory of quantized fields, Lorentz- and
PCT-symmetries are implemented by unitary or antiunitary operators
which arise from the Modular Theory due to Tomita and Takesaki
\cite{BW75,BW76,Haa92}. Because of their close relation to the
algebraic structure of the theory,
these so-called {\em modular symmetries} are
of a particular interest for the algebraic approach
to relativistic Quantum Field Theory founded by Haag and Kastler
\cite{Haa92}. However, a result like the Bisognano-Wichmann Theorem
does not exist for this framework, although the modular objects
considered by Bisognano and Wichmann do exist.
There are even counterexamples
\cite{Yng94} (admittedly, there are, by now, no examples which
are Poincar\'e covariant and satisfy the spectrum condition).

On the other hand, there are applications and partial results
for the algebraic framework. The spin-statistics
connection for parabosonic and parafermionic
massive particle sectors in the sense of Buchholz and
Fredenhagen \cite{BF82}
has recently been derived by different authors from different modular
symmetry assumptions \cite{Ku94a,GL94} (see also \cite{Kuc94,GL95});
the arguments also
work in lower dimensions since any use of the spinor calculus is
avoided. In \cite{GL94}, modular \pct-symmetry is, in addition,
derived from
modular $\upg$-symmetry. For conformal theories of local observables,
the Bisognano-Wichmann modular symmetries have been established
by different groups in different ways
\cite{BGL93,GF93,Joe95}. Conversely, chiral theories in 1+1 dimensions
can even be {\em constructed} by means of their modular objects
\cite{Wie93}.

These results partly depend on a theorem recently found by Borchers
(Theorem II.9 in \cite{Bor92}).
The main implication of this theorem is that in (the vacuum sector of)
a theory satisfying translation covariance and the spectrum
condition,
the Bisognano-Wichmann modular objects commute with
the translation operators in the same
way Lorentz boosts and \pct-operators would commute with
these operators. In 1+1 dimensions, Borchers has concluded
that each local net of observables satisfying translation
covariance and the spectrum condition may be enlarged to a
local net of observables satisfying Poincar\'e convariance
and the Bisognano-Wichmann modular symmetry principles.
The situation in higher dimensions, however, remained an open
problem.

In this paper, we derive from Borchers' commutation relations
a uniqueness result which also holds in higher dimensions:
Borchers' commutation relations imply that the
symmetries which {\em can} be implemented by the Bisognano-Wichmann
modular objects coincide up to at most a translation
with the symmetries obtained in the Wightman setting.
As a symmetry, we admit any unitary or antiunitary operator
under whose adjoint action every algebra of local observables is
mapped onto some algebra which may be localized in some open
region in Minkowski space. Our results are complementary
to those due to Keyl \cite{Key93} and Araki \cite{Ara92}:
in both papers, the notion of a symmetry is more restrictive than
ours\footnote{
   Our notion of a symmetry (which is made precise below) is a very
   general one; a symmetry according to our definition
   need not be a physical
   one. In fact, we show that any Bisognano-Wichmann modular object
   which is a symmetry according to our definition is a {\em physical}
   symmetry.
}, whereas both authors avoid the use of the spectrum condition.
Araki assumes, in addition, that algebras of local observables which are
localized in {\em timelike} separated regions are not contained in
each other's commutants; this property is violated by the massless free
field in any even space-time dimension \cite{Buc75,Bu75a,Buc77,Bu77a}.

\section{Preliminaries and Results}

For some integer $s\geq1$,
denote by $\Rd$ the (1+$s$)-dimensional Minkowski space, and let $V_+$
be the forward light cone. The set $\K$ of all {\em double cones},
i.e., the set of all open sets $\O$ of the form
$$\O:=(a+V_+)\cap(b-V_+),\qquad a,b\in\Rd,$$
is a convenient topological base of $\Rd$. Each nonempty double
cone is fixed by two points, its upper and its lower apex, and
the set $\K$ is invariant under the action of the Poincar\'e group.
For any region $M$ in Minkowski-space, we denote by $\K(M)$ the
set of all double cones contained in $M$.

In the sequel, we denote by $(\H_0,\dA)$ a {\em concrete local
net of observables}: $\H_0$ is a Hilbert space, and the net $\dA$
associates with every double cone $\O\in\K$
a (concrete) C$^*$-algebra $\dA(\O)$ consisting
of bounded operators in $\H_0$
and containing the identity operator; this
mapping is assumed to be {\em isotonous}, i.e., if $\oa\subset\ob$,
                                     $\oa,\ob\in\K$,
then $\dA(\oa)\subset\dA(\ob)$, and {\em local}, i.e.,
if $\oa$ and $\ob$ are spacelike separated double cones and
if $A\in\dA(\oa)$, $B\in\dA(\ob)$, then $AB=BA$.
Since $\K$ is a topological base of $\Rd$, we may for any open
set $M\subset\Rd$ consistently define $\dA(M)$ to be the
C$^*$-algebra generated by the C$^*$-algebras $\dA(\O)$,
$\O\in\K(M)$. We call $\aqloc:=\dA(\Rd)$ the
C$^*$-algebra of {\em quasilocal observables}. Note that every
state of the normed, involutive algebra $\aloc:=\bigcup_{\O\in\K}
\dA(\O)$ of {\em local observables} has a unique continuous extension
to a state of $\aqloc$.

For any subset $M$ of $\Rd$, we denote by $M'$ the
{\em spacelike complement} of $M$, i.e., the set of all points in
$\Rd$ which are spacelike with respect to all points of $M$,
and for every algebra $\M$ of bounded operators in some Hilbert space
$\H$, we denote by $\M'$ the algebra of all bounded operators
which commute with all elements of $\M$.
Using this notation, the above locality assumption  reads
$\dA(\O)\subset\dA(\O')'\quad\forall\O\in\K$.

The net $\dA$ is assumed to be covariant under a strongly
continuous, unitary representation $U$ of the translation group,
and $U$ is assumed to satisfy the spectrum condition, i.e., the support
of the joint spectral measure $E$ of the generators of $U$ is
contained in $\Vpq$. $\dA$ is assumed to satisfy
{\em weak additivity}:
$$\left(\bigcup_{a\in\Rd}\dA(\O+a)\right)''=\aqloc''\qquad\forall
\O\in\K,$$
a useful property which is satisfied by all nets arising from
Wightman fields and
for which we do not know any interesting counterexample.

We assume the existence and uniqueness up to a phase of a
unit vector $\gO$ in $\H_0$ which is invariant under $U$ and
{\em cyclic} with respect to the concrete algebra $(\H_0,\aqloc)$,
i.e. $\overline{\aqloc\gO}=\H_0$;
$\gO$ will be called the {\em vacuum vector}. We assume the
identical representation $(\H_0,\id_{\aqloc})$ of $\aqloc$
to be irreducible, i.e.,
$$\aqloc''=\dB(\H_0),$$ which means that the vacuum state
$A\in \dA\mapsto\langle\gO,A\gO\rangle$ is pure.

Denote by $W$ the wedge
$$W:=\{x\in\Rd:\,x_1\geq|x_0|\}.$$
It follows from the Reeh-Schlieder Theorem \cite{Bor68}
that $\gO$ is cyclic with respect to
$(\H_0,\dA(W)'')$,
and using a standard argument (see, e.g., Prop. 2.5.3 in \cite{BraRo}),
one obtains from locality that $\gO$
is also {\em separating} with respect to $(\H_0,\dA(W)'')$, i.e.,
if $A\in\dA(W)''$ and $A\gO=0$, then $A=0$.
Hence, $(\H_0,\dA(W)'',\gO)$ is a standard von Neumann algebra. We
shall denote by $J$ and $\gD$ the modular conjugation and modular
operator, respectively, of this standard von Neumann algebra according to the
Tomita-Takesaki analysis \cite{Tak70,Haa92,BraRo}.

The most general result which has implications for these modular
objects is the following theorem due to Borchers:

\subsection{Theorem (Borchers' commutation relations)}
\label{bor com rel}
\begin{quote}{\it
Let $(\H,\dM,\Psi)$ be a standard von Neumann algebra with modular
conjugation $J_\dM$ and modular operator $\gD_\dM$, and let
$(T(r))_{r\in\reals}$ be a strongly continuous one-parameter group
of unitaries which has a positive generator and
which for each $r\geq0$ satisfies the conditions}
$$\begin{array}{lrcl}
(a) &T(r)\Psi&=&Psi;\\
(b) &T(r)\dM T(-r)&\subset&\dM.
\end{array}$$
{\it Then for each $r\in\reals$,
the following commutation relations hold:}
$$\begin{array}{lrcl}
(i) &J_\dM T(r)J_\dM&=&T(-r);\\

(ii) &\modopt_\dM T(r)\modopmt_\dM&=&T(e^{2\pi t}r)\qquad\forall
    t\in\reals.\\
\end{array}$$
\end{quote}
Together with a standard argument of Tomita-Takesaki Theory,
this implies, in our setting, for each $a\in\Rd$:
$$\begin{array}{lrcl}
(i) &JU(a)J&=&U(ja),\\
(ii) &\modopt U(a)\modopmt&=&U(V(2\pi t)a)\qquad\forall t\in\reals.
\end{array}$$
where $$j(a_0,a_1,a_2,\dots,a_s):=(-a_0,-a_1,a_2,\dots,a_s),\qquad a\in\Rd,$$
and where $V(2\pi t)$ denotes the Lorentz boost by $2\pi t$ in the
0-1-plane.
So $J$ and $\modopt$, $t\in\reals$, commute with the translations
like a \pct-operator and the group of Lorentz boosts in the
0-1-direction, respectively.
In 1+1 dimensions, Borchers has derived from this
that the net of observables may be enlarged to a local net
which generates the same wedge algebras (and, hence,
the same corresponding modular objects) as the original one
and which has the property that $J$ is a \pct-operator
{\em (modular \pct-symmetry)},
whereas
$\modopt$ implements the Lorentz boost by $2\pi t$ for each
$t\in\reals$ {\em (modular Lorentz-symmetry)}. This was a first
extension of the classical result by Bisognano and Wichmann
\cite{BW75,BW76}, who established modular \pct- and Lorentz symmetry
for Wightman fields in arbitrary spacetime dimensions.

In this paper, we show in particular
that $J$ or $\modopt$, $t\in\reals$,
can be shown to be,
respectively, a \pct-operator or, up to a translation, a 0-1-Lorentz
boost as soon as $J$ or $\modopt$ is {\em any} symmetry. For simplicity,
let us first define a {\em symmetry} to be a unitary or an antiunitary
operator $K$ in $\H_0$ such that
for each $\O\in\K$, there is an open set $M_\O\subset\Rd$ with
$$K\dA(\O)K^*=\dA(M_\O).$$
We are ready now to state our main result:

\subsection{Theorem}\label{Hauptsatz}
\begin{quote}{\it
Let $\dA$ be as above, let $K$ be a symmetry,
and let $\kappa:\,\Rd\to\Rd$ be a linear automorphism such that
$$KU(a)K^*=U(\kappa a)\qquad\forall a\in\Rd.$$
Then there is a unique $\xi\in\Rd$ such that}
$$K\dA(\O)K^*=\dA(\kappa\O+\xi)\qquad\forall\O\in\K.$$
\end{quote}
{}From Theorems \ref{bor com rel} and
\ref{Hauptsatz}, we shall deduce in
particular:

\subsection{Proposition}\label{mod sym}
\begin{quote}{\it
(i) If $J$ is a symmetry, then there is a $\iota\in\Rd$ such that
$$J\dA(\O)J=\dA(j\O+\iota)\qquad\forall\O\in\K.$$
In more than 1+1 dimensions, $\iota=0$. The same holds in
1+1 dimensions as soon as wedge duality or
$\P_+^{\uparrow}$-covariance is given.

(ii) If for some $t\in\reals$, $\modopt$ is a symmetry, then there
is a $b_t\in\Rd$ with $W+b_t=W$ and
$$\modopt\dA(\O)\modopmt=\dA(V(2\pi t)\O+b_t).$$

(iii) If for all $t\in\reals$, $\modopt$ is a symmetry,
then there is a $b\in\Rd$ with $W+b=W$ and
$$\modopt\dA(\O)\modopmt=\dA(V(2\pi t)\O+bt).$$

(iv) If in a theory in 1+$s$ $\geq$1+3 dimensions, $\modopt$
is a symmetry for some $t\in\reals$ and if some nontrivial
rotation $r$ with
$rW=W$ is implemented by a unitary $R$ with $R\gO=\gO$,
then the translation $b_t$ defined in (ii) is trivial.

(v) If in a theory in 1+$s$ $\geq$1+3 dimensions,
$\modopt$ is a symmetry for all $t\in\reals$ and in all
Lorentz frames, then the vector $b$ defined in (iii) is trivial.}
\end{quote}
If in $(i)$, the translation $\iota$ is trivial, it follows from the
results in \cite{GL92} that $J$ is indeed not only a P$_1$T-symmetry,
but even a \pct-operator.

Conversely, $\upg$-covariance is a sufficient condition for the
existence of the rotation $R$. Since it has been shown in \cite{BGL94} that
the assumption made in (v) implies $\upg$-covariance, (v) is an
immediate consequence of (iv).

Proposition\ref{mod sym} is not the only application of Theorem
\ref{Hauptsatz}:

\subsection{Proposition}\label{mod sym+}
\begin{quote}
{\it
Assume $(\H_0,\dA)$ to be Poincar\'e covariant, and assume
that the vacuum vector $\gO$ is not only cyclic, but also
separating with respect to the algebra $(\H_0,\dA(V_+)'')$.

(i) If the modular conjugation $J_+$ of $(\H_0,\dA(V_+)'',\gO)$
is a symmetry, we have
$$J_+\dA(\O)J_+=\dA(-\O)\qquad\forall\O\in\K.$$

(ii) If for some $t\in\reals$, the modular unitary
$\modopt_+$ of $(\H_0,\dA(V_+)'',\gO)$ is a symmetry, we have
$$\modopt_+\dA(\O)\modopmt_+=\dA(e^{-2\pi t}\O).$$}
\end{quote}
Using the scattering theory for massless fermions and bosons in 1+3
or 1+1 dimensions due to
Buchholz \cite{Buc75,Bu75a,Buc77}, it has been shown by Buchholz and
Fredenhagen \cite{Bu75a,Bu77a,BF77} that
each of the modular symmetries considered in this proposition
implies a massless theory to be free (i.e., its S-matrix is trivial).

For the sake of simplicity, we have stated
Theorem \ref{Hauptsatz} and the Propositions \ref{mod sym}
and \ref{mod sym+}
in a form which is not yet the most general one: the notion of a
symmetry may be generalized considerably.
We shall prove our results in the more general form.
As a symmetry, we admit any
unitary or antiunitary operator $K$ in $\H_0$
whose adjoint action preserves
at least fragments of the net structure and therefore carries
localizable algebras to localizable algebras in the following sense:

\subsection{Definition}\label{symmetry}
\begin{quote}
Let $\beta$ be some set of subalgebras of $\dB(\H_0)$ which
contains all local algebras $\dA(\O)$, $\O\in\K$,
as elements, and assume
that with respect to the partial ordering on $\beta$
given by set-theoretic inclusion, all $\B\in\beta$
satisfy $\B=\sup_{\O\in\K_\B}\dA(\O),$
where $\K_\B:=\{\O\in\K:\,\dA(\O)\subset\B\}.$

For any $\B\in\beta$, its {\em localization region}
is given by
$$L(\B):=\bigcup_{\O\in\K_\B}\O.$$

A {\em symmetry} of $(\H_0,\dA)$ with respect to $\beta$
-- or simply: a $\beta$-{\em symmetry} -- is a unitary or
an antiunitary operator $K$ in $\H_0$ with
the property that for any $\O\in\K$, the algebras $K\dA(\O)K^*$
and $K^*\dA(\O)K$ are elements of $\beta$.
\end{quote}
Theorem \ref{Hauptsatz} and Propositions \ref{mod sym}
and \ref{mod sym+} still hold if the term
``symmetry'' is replaced by ``$\beta$-symmetry''. The above formulations
are not the only
interesting applications: if the local algebras are von Neumann algebras,
a possible choice of $\beta$ is the set of von Neumann algebras
generated by local algebras. Furthermore, we need not assume for the
elements $\B\in\beta$ that
$\K_\B=\K(L(\B))$, i.e., not all local algebras localized in the
localization region of $\B$ need to be contained in $\B$;
$L(\B_1)\subset L(\B_2)$ need not imply $\B_1\subset\B_2$.

Before we turn to the proofs of our statements, let us recall a
result which relies on arguments due to Jost,
Lehmann and Dyson \cite{JL57,Dys58} and, in the formulation given
here, on results due to Araki
\cite{Ara63}. See also \cite{Kuc94} for a self-contained exposition
of the proof, and see \cite{BMS61} for a slight generalization.

\subsection{Theorem (Jost, Lehmann, Dyson, Araki)}\label{JLD}
\begin{quote}{\it
Let $K_1$ and $K_2$ be compact subsets of $\Rd$, and let $\psia$ and
$\psib$ be vectors contained in $E(K_1)\H_0$ and $E(K_2)\H_0$,
respectively. Let $A$ and $B$ be local observables, and assume that
the commutator function
$$f_{\psia,\psib,A,B}:\,\Rd\to\complex;\quad x\mapsto\langle
\psia,[A,U(x)BU(-x)]\psib\rangle$$
vanishes in a {\em Jost-Lehmann-Dyson region}, i.e., in an open
subset $R$ of $\Rd$ such that each maximal causal curve in $\Rd$
intersects the region $\overline{R}\cup\overline{R'}$\quad\footnote{This
assumption is avoided in \cite{BMS61}.} and such that two continuous
functions $r_+:\,\reals^s\to\reals$ and
$r_-:\,\reals^s\to\reals$ can be found with the properties:
\begin{quote}
(i) $R = \{x=:(x_0,\xv)\in\Rd:\,r_-(\xv)<x_0<r_+(\xv)\};$

(ii)  $|r_+(\xv)-r_+(\yv)|\leq\|\xv-\yv\|_2\quad\forall\xv,\yv\in\reals^s$;

(iii) $|r_-(\xv)-r_-(\yv)|\leq\|\xv-\yv\|_2\quad\forall\xv,\yv\in
\reals^s$.
\end{quote}
Then the support of $\fpsi$ is contained not only in the complement of
$R$,
but even in the (in general, smaller) union of all {\em admissible
hyperboloids}\footnote{in 1+1 dimensions: admissible hyperbolae},
i.e., the hyperboloids $H_{a,\gs}:=\{x\in\Rd:\,(x-a)^2=\gs^2\}$,
$a\in\Rd$, $\gs\in\reals$, which do not intersect $R$.}
\end{quote}
Another result we shall make use of follows from the Asgeirsson-Lemma,
which describes the behaviour of solutions of the wave equation. For
a proof, see \cite{Ara63}:

\subsection{Lemma (Asgeirsson, Araki)}\label{Asg}
\begin{quote}{\it
If the commutator function $\fpsi$ defined in
Theorem \ref{JLD} and its partial derivatives vanish along a
timelike curve segment $\gg$, they also vanish in the entire double cone
$\gg''$.}
\end{quote}
Note that both these results hold in all dimensions 1+$s$ $\geq$1+1 and they
also apply to the restriction of $\fpsi$ to some timelike plane
(a plane which is invariant under translations in some timelike
direction).

\section{Proofs}

We first show that in 1+s$\geq$1+2 dimensions,
$L(\dA(\O))=\O$ for every double cone $\O\in\K$.
This immediately follows from a Theorem due to Landau
\cite{Lan69}. The identity to be established is
satisfied by any additive theory in 1+s$\geq$1+2 dimensions,
it need not hold in (1+1)-dimensional theories;
the chiral theories are well-known counterexamples. For the proof
of Proposition \ref{mod sym} (i), we shall
need the following, slightly strengthened version of Landau's result:

\subsection{Theorem}\label{Landau}
\begin{quote}{\it
Let $\O$ be a double cone whose closure does not intersect the
closure of $W$. If 1+$s$ $\geq$1+2 or if $\O$ is spacelike with
respect to $W$, we have }
$$\dA(\O')'\cap\dA(W')'=\complex\,\id.$$
\end{quote}
{\bf Proof:}
Let $\C$ be a convex open set such that $\ot:=\O-\C$ and
$\tilde W:=W-\C$ still have disjoint or spacelike separated
closures, respectively. Pick $A\in\dA(\O')'\cap\dA(W')'$ and
$B\in\dA(\C)$, and let $\psia$, $\psib$ and $\fpsi$ be as in
Theorem \ref{JLD}. It follows from locality that $\fpsi$
vanishes in the region $\ot'\cup\tilde W'$.

This region does not admit any
admissible hyperboloid since for any two points
$x$ and $y$ in $\Rd$, the set $(x+\Vq)\cap(y+\Vq)$ \footnote{$\overline{V}$
denotes the closure of the light cone $V:=\{x\in\Rd:\,x^2>0\}$.} contains
some full mass hyperboloid if and only if $x=y$ and since
$$\Rd\backslash(\ot'\cup \tilde W')=
(\ot+\Vq)\cap(\tilde W+\Vq).$$
One verifies that, since $\ot$ and $\tilde W$ are disjoint,
$\ot'\cup \tilde W'$ is a Jost-Lehmann-Dyson
region if and only if $\ot$ is spacelike with respect to
$\tilde W$. In this case, it follows from Theorem \ref{JLD} that
$\fpsi\equiv 0$. Since $A$ and $B$ are bounded and
the vectors with compact momentum space support are
dense in $\H_0$, one obtains $[A,U(x)BU(-x)]=0$ $\forall x\in\Rd$.
Since $B\in\dA(\C)$ is arbitrary, weak additivity gives
$$A\in\left(\bigcup_{x\in\Rd}\dA(\C+x)\right)'=\aqloc'=\dB(\H_0)'
=\complex\,\id,$$
as stated.

Assume now that $\O$ is not necessarily spacelike with respect to
$W$ and that 1+$s$ $\geq$1+2. Sufficiently large shifts $Q+a$
of the plane $Q:=\{x\in\Rd:\,x_2=\dots=x_s=0\}$ have the property
that, with respect to the causal structure the manifold $Q+a$
inherits from Minkowski space, the region $(Q+a)\cap(\ot'\cup\tilde
W')$ is a Jost-Lehmann-Dyson region. In particular, the set
$\alpha$ of all $a\in\Rd$ such that $Q+a$ is of this kind, contains an
open subset of $\Rd$. One verifies that for any shift $Q+a$ of $Q$,
the region $(Q+a)\cap(\ot'\cup\tilde W')$ does not admit any
admissible hyperbola. Theorem \ref{JLD} now implies for every
$a\in\alpha$ that $\fpsi|_{Q+a}\equiv0$. Since $\alpha$ contains
an open region,
we conclude that $\fpsi$ vanishes in a neighborhood of $Q+a$ for
some $a\in\alpha$.
Lemma \ref{Asg} now implies that $\fpsi\equiv0$ all over $\Rd$, from
which $A\in\complex\id$ is obtained as above.
\Halmos

\bigskip\bigskip\noindent%
The theorem is stated in Poincar\'e invariant terms,
so $W$ and $\O$ may be replaced by their respective images under any
Poincar\'e transform.
Note furthermore that the proof still works if $\O$ is replaced by a
causally complete spacelike cone in the sense of \cite{BF82}.
The following lemma is taken from \cite{Ban87}.

\subsection{Lemma}
\begin{quote}{\it
Assume that 1+s$\geq$1+2.
For any two double cones $\oa,\ob\in\K$, $\dA(\oa)$ is a
subalgebra of $\dA(\ob)$ if and only if $\oa\subset\ob$.}
\end{quote}
{\bf Proof:} Because of isotony, we only
need to show that the condition is necessary. If $\oa$ is not
contained in $\ob$ as a subset, there is a double cone $\O$
that is contained in $\oa$ and whose closure does not
intersect the closure of $\ob$. Theorem \ref{Landau} implies
that $\dA(\ob)\cap\dA(\O)=\complex\,\id$. It follows from
additivity that $\dA(\O)\not=\complex\,\id$,
so $\dA(\O)$ is not a subset of $\dA(\ob)$. Since
$\dA(\O)\subset\dA(\oa)$ follows from isotony, $\dA(\oa)$
cannot be a subset of $\dA(\ob)$. $\qquad$
\hfill\Halmos

\bigskip\bigskip\noindent%
This implies that in 1+2 or more dimnensions,
$L(\dA(\O))=\O$ for every double cone $\O\in\K$.

\subsection{Proof of Theorem 1.2}
In the sequel, $\beta$, $K$ and $\kappa$ are defined as in
Definition\ref{symmetry} and Theorem
\ref{Hauptsatz}. We subdivide the proof into a couple of Lemmas.

\subsubsection{Lemma} \label{difference sets}
\begin{quote}
{\it For any algebra $\B\in\beta$, we have
$$ L(K\B K^*)-L(K\B K^*)=\kappa(L(\B)-L(\B)).$$
}
\end{quote}
{\bf Proof:} We have for any $\B\in\beta$:
\begin{eqnarray*}
L(\B)-L(\B)&=&\{a\in\Rd:\,\exists x\in L(\B):\,x+a\in L(\B)\}\\
&=&\{a\in\Rd:\,\exists\O\in\K(L(\B)):\,\O+a\subset L(\B)\}\\
&=&\{a\in\Rd:\,\exists\O\in\K(L(\B)):\,U(a)\dA(\O)U(-a)\subset\B\}\\
&=&\{a\in\Rd:\,\exists\O\in\K_\B:\,U(a)\dA(\O)U(-a)\subset\B\}.
\end{eqnarray*}
Using this and the assumptions of Theorem \ref{Hauptsatz}, we
conclude
\begin{eqnarray*}
& &L(K\B K^*)-L(K\B K^*)\,=\,\{a\in\Rd:\,\exists\O\in\K_{K\B K^*}:\,
U(a)\dA(\O)U(-a)\subset K\B K^*\}\\
& &\hspace{2cm}=\,\{a\in\Rd:\,\exists\O\in\K_{K\B K^*}:\,
KU(\kappa^{-1}a)K^*
\dA(\O)KU(-\kappa^{-1}a)K^*\subset K\B K^*\}\\
& &\hspace{2cm}
=\,\kappa\{a\in\Rd:\,\exists\O\in\K_{K\B K^*}:\,U(a)K^*\dA(\O)KU(-a)
\subset\B\}\\
& &\hspace{2cm}\subset\kappa\{a\in\Rd:\,\exists\O\in\K_{K\B K^*}:\,\exists\ot
\in\K_{K^*\dA(\O)K}:\,U(a)\dA(\ot)U(-a)\subset\B\}\\
& &\hspace{2cm}\subset\,
\kappa\{a\in\Rd:\,\exists\ot\in\K_\B:\,U(a)\dA(\ot)U(-a)\subset\B\}\\
& &\hspace{2cm}=\,\kappa(L(\B)-L(\B)).
\end{eqnarray*}
If we
replace $\B$ by $K^*\B K$ and $K$ by $K^*$, the same reasoning
yields the converse inclusion:
\begin{eqnarray*}
\kappa(L(\B)-L(\B))&=&\kappa(L(K^*K\B K^*K)-L(K^*K\B K^*K))\\
&\subset&\kappa(\kappa^{-1}(L(K\B K^*)-L(K\B K^*))\\
&=&L(K\B K^*)-L(K\B K^*).
\end{eqnarray*}
This proves the Lemma.\Halmos

\subsubsection{Lemma}\label{inclusions}
\begin{quote}{\it For any two double cones $\oa,\ob\in\K$
with $\overline{\oa}\subset\ob$, we have
$$\overline{L(K\dA(\oa)K^*)}\subset L(K\dA(\ob)K^*).$$
}
\end{quote}
{\bf Proof:} $\overline{\oa}\subset\ob$ if and only if the set
$$\{a\in\Rd:\,\oa+a\subset\ob\}$$
is an open neighbourhood of the origin of $\Rd$. Since
\begin{eqnarray*}
\{a\in\Rd:\,\oa+a\subset\ob\}&=&\{a\in\Rd:\,U(a)\dA(\oa)U(-a)
\subset\dA(\ob)\\
&=&\{a\in\Rd:\,K^*U(\kappa^{-1} a)K\dA(\oa)K^*U(-\kappa^{-1} a)K
\subset \dA(\ob)\}\\
&=&\{a\in\Rd:\,U(\kappa^{-1}a)K\dA(\oa)K^*U(-\kappa^{-1}a)
\subset K\dA(\ob)K^*\}\\
&=&\kappa\{a\in\Rd:\,U(a)K\dA(\oa)K^*U(-a)
\subset K\dA(\ob)K^*\}\\
&\subset&\kappa\{a\in\Rd:\,L(K\dA(\oa)K^*)+a\subset
L(K\dA(\ob)K^*)\},
\end{eqnarray*}
and since $\kappa$ is a homeomorphism of $\Rd$ onto itself, it
follows that this set is an open neighbourhood of the origin if and only if
$$\{a\in\Rd:\,L(K\dA(\oa)K^*)+a\subset L(K\dA(\ob)K^*)\}$$
is an open neighbourhood of the origin. This gives the statement.
\halmos

\subsubsection{Lemma}\label{bases}
\begin{quote}{\it
Let $x\in\Rd$ be arbitrary, and
let $(\onu)_{\nu\in\naturals}$ be a neighbourhood base of
$x$ consisting of double cones $\onu\in\K$.

Then $(L(K\dA(\onu)K^*))_{\nu\in\naturals}$ is a neighbourhood base of a
(naturally, unique) point $\tilde\kappa(x)\in\Rd$.}
\end{quote}
{\bf Proof:}
Without loss of generality, we may assume that
$\overline{\O_{\nu+1}}\subset\onu
\quad\forall\nu\in\naturals$. It follows from
$L(\dA(\O))=\O\quad\forall\O\in\K$ and Lemma \ref{difference sets}
that all $L(K\dA(\onu)K^*)$, $\nu\in\naturals$, are bounded sets,
and it follows from Lemma \ref{inclusions} that
$$\overline{L(K\dA(\O_{\nu+1})K^*)}\subset L(K\dA(\onu)K^*).$$
Therefore, the intersection of this family is nonempty,
and Lemma \ref{difference sets} implies that the
diameter of $L(K\dA(\onu)K^*)$ tends to zero as $\nu$ tends to infinity.
This implies that the intersection contains precisely one point
$\tilde\kappa(x)$,
as stated.\halmos

\subsubsection{Corollary}
\begin{quote}{\it
The map $\tilde\kappa$ given by Lemma \ref{bases}
is a homeomorphism, and for any algebra
$\B\in\beta$, we have}
$$L(K\B K^*)=\tilde\kappa(L(\B)).$$
\end{quote}

\subsubsection{Lemma}
\begin{quote}{\it
For each $x\in\Rd$, we have
$$\tilde\kappa(x)=\tilde\kappa(o)+\kappa x,$$
where $o$ denotes the origin of $\Rd$.}
\end{quote}
{\bf Proof:} Let $(\onu)_{\nu\in\naturals}$ be a neighborhood
base of $o$. Then $(\onu+x)_{\nu\in\naturals}$ is a neighborhood
base of $x$, and
$$\bigcap_{\nu\in\naturals}L(K\dA(\onu+x)K^*)=\bigcap_{\nu\in\naturals}
\tilde\kappa(\onu+x)=\{\tilde\kappa(x)\}.$$
On the other hand, we have
\begin{eqnarray*}
\bigcap_{\nu\in\naturals}L(K\dA(\onu+x)K^*)&=&\bigcap_{\nu\in\naturals}
L(U(\kappa x)K\dA(\onu)K^*U(-\kappa x))\\
&=&\kappa x+\bigcap_{\nu\in\naturals}\tilde\kappa(\onu)\\
&=&\kappa x+\{\tilde\kappa(o)\}.
\end{eqnarray*}\halmos

\bigskip\bigskip\noindent%
This proves Theorem \ref{Hauptsatz}, where $k=\tilde\kappa(o)$.

\subsection{Proof of the Propositions 1.3 and 1.4}

To start with (i) of Proposition 1.3,
it follows from Theorem \ref{Hauptsatz} that
there is a homeomorphism $\jt$ of $\Rd$ into itself such that
$$J\dA(\O)J=\dA(\jt(\O)) \qquad\forall \O\in\K.$$
This $\jt$ satisfies the relation
$$\jt(x)=\jt(o)+jx=:\iota+j x\qquad\forall x\in\Rd.$$
We have to show that $\iota=o$.
Since $J$ is an involution, so is $\jt$. This implies
$$x=\jt(\jt(x))=\jt(\iota+j x)=\iota+j \iota+x\qquad\forall
x\in\Rd,$$
which gives $\iota=-j\iota$, hence $\iota_2=\dots=\iota_s=0$.
It remains to show that $\iota_0=\iota_1=0$, which is
equivalent to $\jt(W)=W'$. To this end, note that $W'\subset
\jt(W)$ already follows from locality.
If wedge duality is assumed, the converse inclusion, which
completes the proof, is trivial, and
if, instead, $\P_+^{\uparrow}$-covariance is assumed, the statement
immediately
follows from the fact that the 0-1-boosts commute with $J$ and,
therefore, leave the algebra $J\dA(W)''J=\dA(\jt(W))''$
invariant.

Now let us just assume that 1+s$\geq$1+2 and that
$\jt(W)\not\subset W'$. Then there is a double cone $\O\in
\K(\jt(W))$ whose closure is disjoint from $\overline W'$.
Because of
$$\K(\jt(W))=\K_{J\dA(W)''J}=\K_{\dA(W)'},$$
we conclude $\dA(\O)\subset\dA(W)'$. Theorem \ref{Landau}
now implies that $\dA(\O)=\complex\,\id$, and by additivity,
the whole net has to be trivial. This, however, is in conflict
with irreducibility, which implies $\jt(W)=W'$.

(ii) and (iii) are immediate consequences of Theorem
\ref{Hauptsatz}.

(iv) follows from Theorem \ref{Landau} since for each $\O\in\K$,
we have, on the one hand:
$$R\modopt\dA(\O)\modopmt R^*=R\dA(V(2\pi t\O+b_t)R^*=
\dA(rV(2\pi t)\O+rb_t),$$
whereas on the other hand, $R\dA(W)''R^*=\dA(W)''$ and $R\gO=\gO$
imply $R\modopt=\modopt R$ so that
$$R\modopt\dA(\O)\modopmt R^*=\modopt\dA(r\O)\modopmt
=\dA(V(2\pi t)r\O+b_t))=\dA(rV(2\pi t)\O+b_t);$$
comparing the two expressions and using
$L(\dA(\O))=\O\quad\forall\O\in\K$ yields the stated result.

(v) has been derived from (iv) in the remark following
Proposition \ref{mod sym}.

To prove Proposition 1.4, note that Theorem \ref{bor com rel}
implies for all $a\in\Rd$ the commutation relations
\begin{eqnarray*}
J_+U(a)J_+&=&U(-a);\\
\modopt_+U(a)\modopmt_+&=&U(e^{2\pi t}a)\qquad\forall t\in\reals.
\end{eqnarray*}
If, respectively
$J_+$ or $\modopt_+$ is a symmetry, Theorem \ref{Hauptsatz} implies
that it can differ from the stated symmetry at most by a translation.
Since $V_+$ is Lorentz-invariant, the modular data of
$(\H_0,\dA(V_+)'',\gO)$ commute with all $U(g),\,g\in\poin.$
However, there are no nontrivial translations which commute
with all $g\in\poin$; this proves Proposition 1.4.

\end{document}